\documentclass
[superscriptaddress,secnumarabic,amssymb,amsmath,nobibnotes,aps,prd,showkeys,nofootinbib,nopacsnumber,onecolumn,10pt]{revtex4}%
\usepackage{graphicx}
\usepackage{bm}
\usepackage{xcolor}
\usepackage{amsmath}
\usepackage{amsfonts}
\usepackage[english]{babel}
\usepackage{amssymb}
\usepackage[utf8]{inputenc}%
\providecommand{\U}[1]{\protect\rule{.1in}{.1in}}
\newcommand{\be}{\begin{equation}}
\newcommand{\ee}{\end{equation}}

\newcommand{\mincir}{\raise
-3.truept\hbox{\rlap{\hbox{$\sim$}}\raise4.truept\hbox{$<$}\ }}
\newcommand{\magcir}{\raise
-3.truept\hbox{\rlap{\hbox{$\sim$}}\raise4.truept\hbox{$>$}\ }}

\begin{document}
\title{Testing the Coexistence of Dark Energy and Dark Matter with Late-time Observational Data}
\author{Shambel Sahlu}
\email{shambel.sahlu@nithecs.ac.za}
\affiliation{Centre for Space Research, North-West University, Potchefstroom 2520, South Africa}
\affiliation{National Institute for Theoretical and Computational Sciences (NITheCS), South Africa.}
\author{Andronikos Paliathanasis}
\email{anpaliat@phys.uoa.gr}
\affiliation{Centre for Space Research, North-West University, Potchefstroom 2520, South Africa}
\affiliation{Institute of Systems Science, Durban University of Technology, Durban 4000,
South Africa}
\affiliation{Departamento de Matem\`{a}ticas, Universidad Cat\`{o}lica del Norte, Avda.
Angamos 0610, Casilla 1280 Antofagasta, Chile}
\affiliation{National Institute for Theoretical and Computational Sciences (NITheCS), South Africa.}
\author{Genly Leon}
\email{genly.leon@ucn.cl}
\affiliation{Departamento de Matem\`{a}ticas, Universidad Cat\`{o}lica del Norte, Avda.
Angamos 0610, Casilla 1280 Antofagasta, Chile}
\affiliation{Institute of Systems Science, Durban University of Technology, Durban 4000,
South Africa}
\affiliation{Centre for Space Research, North-West University, Potchefstroom 2520, South Africa}
\author{Amare Abebe}
\email{amare.abebe@nithecs.ac.za}
\affiliation{Centre for Space Research, North-West University, Potchefstroom 2520, South Africa}
\affiliation{National Institute for Theoretical and Computational Sciences (NITheCS), South Africa.}

\begin{abstract}
We investigate the viability of a cosmological scenario with interacting dark sector, which can describe the coexistence between dark energy and
dark matter. The model possesses an analytical solution for the Hubble function
and we constrain the free parameters by applying the newly released cosmic
chronometers data (31 old data and 3 new data from DESI), the Baryonic
Acoustic Oscillators from the Dark Energy Spectroscopic Instrument Survey
(DESI DR2 BAO), along with Gamma-ray bursts (GRBs) and Supernova catalogues
(Pantheon Plus, Union3, and DES-Dovekie). We find that the coexistence model
fits the data sets in a better way than the reference models - the
$\Lambda$CDM and $w$CDM models. The analysis shows that the coexistence scenario can provide a cosmologically viable model for the
description of the late-time acceleration of the universe. Nevertheless, for
large redshifts, the model has a similar behaviour to that of the $w$CDM
model, as the introduction of the GRB data indicates in the statistical
parameters. Finally, it is worth mentioning that the coexistence model
provides a statistically smaller value for the $H_{0}$ parameter.

\end{abstract}
\date{\today}
\maketitle

\section{Introduction}

The cosmological observations indicate that the universe is currently
undergoing an accelerated expansion
\cite{riess1998observational,pope2004cosmological,spergel2003first}. According
to General Relativity, this behaviour requires the introduction of an exotic
component, known as dark energy, into Einstein's field equations, a form of
energy with negative pressure capable of driving the cosmic acceleration. The
origin and nature of dark energy remain one of the challenging problems in
modern cosmology \cite{Brax:2017idh,Huterer:2017buf}.

The cosmological constant, which gives rise to the $\Lambda$CDM model, remains
the simplest dark-energy candidate. Despite its well-known theoretical
pathologies \cite{Perivolaropoulos:2021jda} and the persistent cosmological
tensions \cite{CosmoVerseNetwork:2025alb}, it provides an excellent
description of current observations. Nevertheless, the 2024 release of the
Baryon Acoustic Oscillation (BAO) measurements from the Dark Energy
Spectroscopic Instrument (DESI DR1) has challenged the long-standing status of
$\Lambda$CDM among dark energy models \cite{DESI:2024mwx}. Moreover, the subsequent BAO data of the DESI DR2 release
\cite{DESI:2025zpo,DESI:2025zgx,DESI:2025fii}, published in 2025, have provided
further support to the dynamical character of dark energy, with a possible
scenario for the equation of state parameter for the dark energy to cross the
phantom divide line.

There are various studies in the literature that explore the new cosmological
data, dark energy models with a parametric equation of state parameter were
introduced in \cite{Ormondroyd:2025iaf,Kessler:2025kju,Mirpoorian:2025rfp,
Scherer:2025esj,Nagpal:2025omq,Hussain:2025nqy,Ozulker:2025ehg,
Li:2025ops,Paliathanasis:2025cuc,Malekjani:2025alf,Carloni:2025dqt}. On the
other hand, there are a plethora of studies that provide mechanisms for the
theoretical justification of the dynamical character of dark energy, such as
scalar fields
\cite{Hossain:2025grx,Shajib:2025tpd,Anchordoqui:2025fgz,Efstratiou:2025iqi,
Paliathanasis:2025xxm,Paliathanasis:2025mvy,Alestas:2025syk,SanchezLopez:2025uzw,
Pourtsidou:2025sdd,Samanta:2025oqz,Dinda:2025iaq,Cai:2025mas}, modified
theories of gravity
\cite{Yang:2025mws,Lu:2025sjg,Feng:2025cwi,Karmakar:2025yng,Samaddar:2025bjg,
Philcox:2025faf,Paliathanasis:2025hjw,Odintsov:2025jfq,Plaza:2025gcv,
DOnofrio:2025cuk,Najera:2025htf}, and other approaches, see, for instance, and
reference therein \cite{Ye:2025pem,Toomey:2025yuy,Zheng:2025vyv,
Shen:2025cjm,Hussain:2025uye,Yadav:2025vpx,Paliathanasis:2025dcr,
Paliathanasis:2025kmg,Yadav:2025wbc,Luciano:2025elo,Luongo:2024fww,Alfano:2024fzv,Alfano:2025gie}%
. A special class of models that have drawn attention are the interacting
models \cite{Li:2024qso, Giare:2024smz, Zhu:2025lrk, Zhang:2025dwu,Li:2025muv,
Petri:2025swg, Li:2025ula, Pan:2025qwy, Silva:2025hxw, Shah:2025ayl,
Feng:2025mlo, Yang:2025uyv, Escobal:2026lnp,Paliathanasis:2026ymi,Li:2026xaz}.

Interacting models describe the flow of energy between the constituents of the
dark sector, namely dark matter and dark energy, offer a simple and efficient mechanism for addressing several cosmological problems as well as describing
the dynamical nature of dark energy \cite{Baldi:2008ay,Tocchini-Valentini:2001wmi,
Chimento:2007yt, SantanaJunior:2024cug, Chimento:2012aea, Pan:2012ki,
Sharov:2017iue, Yang:2018ubt, Paliathanasis:2019hbi, Paliathanasis:2024jxo,
Leon:2020pvt, Okengo:2024mub, Valiviita:2008iv, Caldera-Cabral:2008yyo,
vanderWesthuizen:2025vcb,
vanderWesthuizen:2025mnw,vanderWesthuizen:2025rip,Cruickshank:2025chm,Silva:2025bnn}
without a phantom crossing scenario \cite{Guedezounme:2025wav}. Cosmological
interacting models can be viewed as a natural extension of a varying vacuum
scenario \cite{Basilakos:2009ms,Basilakos:2009ah,
Yang:2018uae,Pan:2014lua,Pan:2018ibu,Fritzsch:2015lua,Fritzsch:2016ewd}, since
the latter model is recovered when the dark energy component reduces to the
cosmological constant. A broad discussion of the influence of the interaction
between dark energy and dark matter, and its effect on structure growth and CMB, is presented in \cite{pu2015early,murgia2016constraints}. In \cite{zhai2023consistent} the role of CMB probes in the interaction of dark energy models was emphasized, 
utilizing data from different sources. It was found that interaction 
models affect the value of the Hubble constant, and in
\cite{lucca2021multi,wang2016dark} cosmological interactions have been used as
a mechanism to address the $H_{0}$ tension \cite{DiValentino:2017iww}.
In addition, interacting models have been used to address the $S_{8}$ tension
problem \cite{Shah:2024rme}.

At the fundamental level, the cosmological
medium is a single fluid composed of different components. For a nonzero
interacting term, the cold dark matter no longer behaves as a pressureless
fluid because effective pressure terms are introduced, while the equation of
state parameter of the dark energy sector is also modified. Thus, observable
effects of the interaction are attributed to the dynamical character of the
cosmic fluid.

Recently, in \cite{2025PDU4701750P}, the asymptotic dynamics were revisited
for two interacting models previously proposed in
\cite{2011GReGr..43.1309K,2023Univ....9..437O}, which also share the
characteristic feature that the Hubble function can be expressed in closed form. Furthermore, similarities and analogies with dynamical population 
models of ecological systems were explored, showing that these two models can
describe compartmentalization and coexistence within the dark sector. The
compartmentalization scenario was examined using late-time cosmological
observations in \cite{vanderWesthuizen:2025iam}. It was found that the model
is in good agreement with the cosmological data, particularly when the BAO
measurements are included.

In this work, we investigate whether current observational data for the
late-universe supports the coexistence interacting model and whether it can
challenge the $\Lambda$CDM cosmology. To perform our analysis, we employ the
newly released BAO measurements from the DESI DR2 catalogue, together with the
latest updated Observable Hubble Data (OHD) obtained from Cosmic Chronometers
(CC). Specifically, we use a total of 34 measurements: the 31 data points
described in \cite{moresco2020setting, qi2023model}, along with the three new
determinations extracted from the DESI analysis as presented in
\cite{Loubser:2025fzl}, namely $\left(  z,H(z)\right)  =\left(
0.46,88.48^{\pm0.57\pm12.32}~km/s/\text{Mpc}\right)
$, $\left(  z,H(z)\right)  =\left(  0.67,119.45^{\pm6.39\pm16.64}%
~\text{km}/s/\text{Mpc}\right)  $, and $\left(
z,H(z)\right)  =\left(  0.83,108.28^{\pm10.07\pm15.08}~\text{km}/s/\text{Mpc}\right)  $. Additionally, we consider the PantheonPlus
\cite{Brout:2022vxf}, the Union3.0 \cite{rubin2023union} and DES-Dovekie
\cite{DES:2025sig} supernova catalogues, and 193 gamma-ray-bursts events (GRB)
within the redshift $0.0335<z<8.1~$which have been calibrated using the Amati
correlation \cite{Amati:2018tso}.

The structure of the paper is as follows. In Section 2, we briefly discuss the effects of dark energy--dark matter
interactions in Einstein's General Relativity and present the model under
consideration. The dynamical evolution of the physical variables for this
model is described by a two-dimensional Lotka--Volterra system, which admits
an analytic solution. In Section 3, we test the corresponding Hubble function
against the cosmological data, where we find that the coexistence interacting
model fits the different combinations of the above-described datasets better
than the two reference models, $\Lambda$CDM and $w$CDM. Finally, in
Section 4, we discuss our findings and present our conclusions.

\section{Interacting Dark Sector}

Within the framework of General Relativity, the gravitational field is
described by the action integral
\begin{equation}
S=\int{d^{4}x\sqrt{-g}\left[  \frac{R}{2}+\mathcal{L}_{tot}\right]  }\;,
\end{equation}
where $R$ is the Ricci scalar constructed by the Levi-Civita connection of
the metric tensor $g_{\mu\nu}$, $\mathcal{L}_{tot}~$is the total Lagrangian density
of the cosmological fluid.\ The latter we assume that it describes two main
components, the dark energy $\mathcal{L}_{DE}$, and the cold dark matter
$\mathcal{L}_{m}$, i.e. $\mathcal{L}_{tot}=\mathcal{L}_{DE}+\mathcal{L}_{m}$, which are the major components of the cosmic fluid.

Variation with respect to the metric tensor leads to the gravitational field
equations
\begin{equation}
G^{\mu\nu}\equiv T^{\mu\nu}_{tot}\;, \label{efe}%
\end{equation}
where $T^{\mu\nu}_{\mathrm{tot}}=T^{\mu\nu}_{m}+T^{\mu\nu}_{DE}\;$, with $T_{m
}^{\mu\nu}=\frac{1}{\sqrt{-g}}\frac{\partial\left(  \sqrt{-g}\mathcal{L}%
_{m}\right)  }{\partial g^{\mu\nu}}$ and $T^{\mu\nu}_{DE}=\frac{1}{\sqrt{-g}%
}\frac{\partial\left(  \sqrt{-g}\mathcal{L}_{DE}\right)  }{\partial g^{\mu\nu}%
}$. \newline Furthermore, the Bianchi identity reveals the conserved nature of
the gravitational field, that is, $\nabla_{\nu}G^{\mu\nu}=0$ and $\nabla_{\nu
}\left(  T_{\mathrm{tot}}^{~~\;\;\mu\nu}\right)  =0$.

On a very large scale, the universe is considered to be isotropic and homogeneous, described by the four-dimensional spatially flat FLRW geometry
with line element
\[
ds^{2}=-dt^{2}+a^{2}\left(  t\right)  \left(  dx^{2}+dy^{2}+dz^{2}\right)
\text{,}%
\]
where $a(t)$ describes the radius of the three-dimensional hypersurface.

The cosmological fluid is described by a homogeneous perfect fluid, that is,
the energy momentum tensor~$T_{\mu\nu}^{tot}$ reads
\begin{equation}
T^{\mu\nu}_{tot}=(\rho+p)u^{\mu}u^{\nu}+pg^{\mu\nu}\;, \label{emt}%
\end{equation}
where$~u^{\mu}=\delta_{t}^{\mu}$ is the comoving observer and $\rho=\rho(t)$
and $p=p(t)$ are the effective energy density and pressure components of the
cosmic fluid, respectively. Therefore, $\rho=\rho_{m}+\rho_{d}$ and $p=p_{d}$,
with $p_{d}=w_{d}\rho_{d}$. For the comoving observer the expansion rate of
the three-dimensional space is defined as $\theta=3H$, where $H=\frac{\dot{a}%
}{a}$ is the Hubble function.

In the FLRW background, the conservation equation for the cosmological fluid
reads%
\begin{equation}
\dot{\rho}+3H\left(  \rho+p\right)  =0\text{,}%
\end{equation}
or equivalently%
\begin{equation}
\dot{\rho}+3H\left(  1+w_{tot}\right)  \rho=0\text{,}%
\end{equation}
in which $w_{tot}=\frac{p}{\rho}$ is the equation of state parameter for the
cosmic fluid. The latter equation can be expressed in the equivalent form
\begin{align}
\dot{\rho}_{m}+3H\rho_{m}  &  =Q(t),\\
\dot{\rho}_{d}+3H\left(  1+w_{d}\right)  \rho_{d}  &  =-Q(t).
\end{align}
where $Q\left(  t\right)  $ describes the energy transfer between the two fluid sources. When the two fluids are minimally coupled, then $Q\left(
t\right)  =0$, and for a constant equation of state parameter $w_{d}$ for the dark energy fluid, the latter system provides the closed-form solution
\[
\rho_{m}=\rho_{m0}a^{-3},~\rho_{d}=\rho_{d0}a^{-3\left(  1+w_{d}\right)  },
\]
in which $\rho_{m0}$, and $\rho_{d0}$ describe the energy density for the two
fluids at the present.

Nevertheless, when the two fluids interact, the latter solution is not valid.
The energy transfer between the two fluids leads to the introduction of
effective pressure components for the individual fluids, leading to a new
dynamical behaviour for the total fluid.

In this study, we consider the following interaction \cite{2025PDU4701750P}%
\[
Q\left(  t\right)  =Q_{0}\left(  t\right)  \rho_{m}\rho_{d}+Q_{1}\left(
t\right)  \rho_{d},
\]
where~$Q_{0}\left(  t\right)  =\frac{\alpha}{H}$ and $Q_{1}\left(  t\right)
=\beta H$, with $\alpha$,$~\beta$ to be coupling constants. This interacting
model has the characteristics of the Lotka-Volterra system and leads to a cosmological model that can describe a dark energy-dark matter coexistence
solution. In the limit where $\beta=0$, then the model which describes
compartmentalization in the dark sector is recovered \cite{2025PDU4701750P}.

The analytical solution and the cosmological dynamics for the field equations for this Lotka-Voltera model has been investigated in detail before in
\cite{2025PDU4701750P}. The closed-form solution reads
\begin{equation}
H\left(  a\right)  =H_{0}\left(  1-\hat{\Omega}_{m0}+\hat{\Omega}_{m0}%
a^{p_{1}}\right)  ^{p_{2}}a^{-3\left(  1+p_{3}\right)  }, \label{sc.01}%
\end{equation}
where $p_{1}$,~$p_{2}$ and $p_{3}$ are%
\begin{align}
p_{1}  &  =3\left(  \alpha+w_{d}\right)  +\beta\,,\\
p_{2}  &  =\frac{w_{d}}{2\left(  \alpha+w_{d}\right)  },\\
p_{3}  &  =3w_{d}\left(  1+\frac{\beta}{3\left(  \alpha+w_{d}\right)
}\right)  .
\end{align}
Parameter $\hat{\Omega}_{m0}$ is an integration constant, and it should not be
confused with the energy density of the dark matter. Only
when $\alpha=\beta=0$, the analytic solution for the $w$CDM model
is recovered, and that of the $\Lambda$CDM for $w_{d}=-1$, then $\hat{\Omega
}_{m0}=\Omega_{m0}=\frac{\rho_{m0}}{3H_{0}^{2}}$. \ 

It is important to note that in the interacting scenario, the two fluids cannot
be treated as distinct components. In reality, there is a single cosmic fluid
that we interpret as having two effective components.

\section{Observational Tests}

In this Section, we present the observational tests of our model.
Specifically, we aim to investigate whether the dark matter--dark energy
coexistence scenario is viable and whether it is supported by cosmological data.

\subsection{Observational Data \& Methodology}

\label{Obs}

For the observational tests, we consider data given by the Supernova
catalogues, the Cosmic Chronometer, which provide Observational Hubble data,
Baryonic Acoustic Oscillators and Gamma Ray Bursts. In particular:

\begin{itemize}
\item[-] (PP/U3/DESD) We consider three different sets of data for the
Supernova (SNIa), the PantheonPlus (PP) \cite{Brout:2022vxf} the Union3.0
\cite{rubin2023union} and the recent DES-Dovekie (DESD) catalogues. The PP
dataset contains 1,701 light curves corresponding to 1,550 spectroscopically
confirmed supernovae, and here we consider it without the SH0ES Cepheid
calibration. The U3 catalogue is the most recent compilation, including 2,087
events in the same redshift range as PP, that is, $10^{-3}<z<2.27~$, of which
1,363 are shared with PantheonPlus. The main difference between the two
catalogues lies in the methodology used to analyze the direct photometric
observations. On the other hand, the DESD catalogue follows from a re-analysis
with an improved photometric calibration of the five-year data of the Dark
Energy Survey \ of Type Ia supernova (DES-SN5YR). DESD catalogue includes 1820
SNIa events with the redshifts $z<1.13$ \cite{DES:2025sig}. For the spatially
flat FLRW geometry the theoretical distance modulus $\mu^{th}$ is calculated
from the distance modules, $D_{L}=c\left(  1+z\right)  \int\frac{dz}{H\left(
z\right)  }$, using the relation $\mu^{th}=5\log D_{L}+25$. Both datasets
provide measurements of the apparent magnitude $m~$and calibrated values of
the distance modulus $\mu^{obs}$.

\item[-] (OHD) Cosmic chronometers provide an independent probe of the
expansion history by using passively evolving galaxies with nearly
synchronized stellar populations \cite{moresco2020setting}. In this study, we
consider the 34 direct measurements of the Hubble parameter for redshifts in
the range $0.09\leq z\leq1.965$. The 31 data are provided by the catalogue in
\cite{moresco2020setting} while we consider the three new measurements
from the analysis of the DESI DR1 data \cite{Loubser:2025fzl}.

\item[-] (BAO) We make use of the recent BAO data of DESI DR2 release
\cite{DESI:2025zpo,DESI:2025zgx,DESI:2025fii}. The dataset includes
observation values of the transverse comoving angular distance ratio
$D_{M}r_{drag}^{-1}=\left(  1+z\right)  ^{-1}D_{L}r_{drag}^{-1},$ the volume
averaged distance ratio$~D_{V}r_{drag}^{-1}=\left(  cD_{L}zH\left(  z\right)
^{-1}\right)  ^{\frac{1}{3}}r_{drag}^{-1}$~and the Hubble distance
ratio$~D_{H}r_{d}^{-1}=cH\left(  z\right)  ^{-1}r_{drag}^{-1}$ where
$r_{drag}$ is the sound horizon at the drag epoch.

\item[-] (GRBs) We consider GRBs which provide us with observational tests for
larger values of redshifts. In particular, we employ the 193 GRBs events in the
redshift range $0.0335<z< 8.1$ analyzed through the Amati correlation
\cite{Amati:2018tso}.
\end{itemize}

For our study, we consider different combinations for the
datasets. Specifically, we consider the following combinations of data:
OHD + BAO; PP + OHD + BAO; U3 + OHD + BAO; DESD + OHD + BAO; OHD + BAO + GRBs;
PP + OHD + BAO + GRBs; U3 + OHD + BAO + GRBs and DESD + OHD + BAO + GRBs. With the
latter, we will be able to understand how our model fits each data set, and
specifically to understand the statistical behaviour of our model on larger
redshifts when the GRB are introduced.

For the statistical analysis, we use the Bayesian inference framework
COBAYA\footnote{https://cobaya.readthedocs.io/}~\cite{cob1,cob2} with the MCMC
sampler~\cite{mcmc1,mcmc2} where the theoretical observational parameters are
calculated with the use of the custom theory. The chains of the MCMC sampler
are studied using the GetDist
library\footnote{https://getdist.readthedocs.io/}~\cite{getd}\thinspace.

For the Coexistence interacting model, the parametric space has dimensions six,
consisting of the free parameters $\left\{  H_{0},\hat{\Omega}_{m0}%
,r_{drag},w_{d},\alpha,\beta\right\}  $. Furthermore, we compare our model
with the $\Lambda$CDM and the $w$CDM, which are our reference models.
These two models are recovered from the coexistence interacting model when
$\left(  \alpha,\beta\right)  =\left(  0,0\right)  $, and for the $\Lambda
$CDM, $w_{d}=-1$. \ In Table \ref{table0} we present the priors considered in
this work for the study of the model. The contribution of the baryons at the
late time is very small. Thus, we consider the Hubble function (\ref{sc.01})
as a good approximation for the description of the late universe.%

%TCIMACRO{\TeXButton{B}{\begin{table}[tbp] \centering}}%
%BeginExpansion
\begin{table}[tbp] \centering
%EndExpansion
\caption{Priors for the free parameters for the observational tests.}%
\begin{tabular}
[c]{cccc}\hline\hline
& $\mathbf{\Lambda}$\textbf{CDM} & $w$\textbf{CDM} & \textbf{Coexistence}%
\\\hline
$\mathbf{H}_{0}$ & $\left[  60,80\right]  $ & $\left[  60,80\right]  $ &
$\left[  60,80\right]  $\\
$\mathbf{\hat{\Omega}}_{m0}$ & $\left[  10^{-3},0.5\right]  $ & $\left[
10^{-3},0.5\right]  $ & $\left[  10^{-3},0.5\right]  $\\
$\mathbf{r}_{drag}$ & $\left[  130,170\right]  $ & $\left[  130,170\right]  $
& $\left[  130,170\right]  $\\
$\mathbf{w}_{d}$ & $-1$ & $\left[  -1.2,-0.33\right]  $ & $\left[
-1.2,-0.33\right]  $\\
$\alpha$ & $0$ & $0$ & $\left[  -2,2\right]  $\\
$\beta$ & $0$ & $0$ & $\left[  -1,1\right]  $\\\hline\hline
\end{tabular}
\label{table0}%
%TCIMACRO{\TeXButton{E}{\end{table}}}%
%BeginExpansion
\end{table}%
%EndExpansion

Because these three models have different dimensions for the parametric space,
we make use of the Akaike Information Criterion (AIC)~\cite{AIC} which allows
us to perform a statistical comparison. The AIC parameter is approximately
defined, for a large number of data, from the $\chi_{\min}^{2}$, from the
relation%
\begin{equation}
{AIC}\simeq\chi_{\min}^{2}+2\kappa,
\end{equation}
where $\kappa=6$ for the coexistence model, $\kappa=3$ for the $\Lambda$CDM
$\ $and $\kappa=4$ for the $w$CDM.

Akaike's scale provides information regarding the preferred model for each
dataset. Specifically, from the difference of the AIC parameters for two
models, $\Delta AIC$, Akaike's scale indicate that when $\lvert\Delta
{AIC}\rvert<2$, the two models are statistically equivalent, for
$2<\lvert\Delta{AIC}\rvert<6$, there is weak evidence in favor of the model
with the smaller AIC parameter, for $6<\lvert\Delta{AIC}\rvert<10$, the
evidence is strong, and for $\lvert\Delta{AIC}\rvert>10$, there is clear
evidence supporting the preference for the model with the lower AIC parameter.

\subsection{Results}

In the following, we present the results from the analysis of the MCMC chains for
the six different combined datasets we discussed above. The results of the
MCMC chains are summarized in\ Tables \ref{table1} and \ref{table2}, while the
comparison of the statistical parameters are given in Tables \ref{table3} and
\ref{table4}.%

%TCIMACRO{\TeXButton{B}{\begin{table}[tbp] \centering}}%
%BeginExpansion
\begin{table}[tbp] \centering
%EndExpansion
\caption{Posterior Parameters for the interacting model which describes coexistence (1/2).}%
\begin{tabular}
[c]{ccccccc}\hline\hline
& $\mathbf{H}_{0}$ & $\mathbf{\hat{\Omega}}_{m0}$ & $\mathbf{r}_{drag}$ &
$\mathbf{w}_{d}$ & $\alpha$ & $\beta$\\\hline
\multicolumn{7}{c}{\textbf{Dataset OHD + BAO}}\\
$\mathbf{\Lambda}$\textbf{CDM} & $69.4_{-1.7}^{+1.7}$ & $0.296_{-0.014}%
^{+0.014}$ & $146.9_{-3.3}^{+3.3}$ & $-1$ & $0$ & $0~$\\
$w$\textbf{CDM} & $68.5_{-2.7}^{+2.4}$ & $0.295_{-0.016}^{+0.016}$ &
$146.9_{-3.4}^{+3.4}$ & $-0.94_{-0.12}^{+0.12}$ & $0$ & $0$\\
\textbf{Coexistence} & $68.1_{-2.8}^{+2.3}$ & $<0.175$ & $146.9_{-3.3}^{+3.3}$
& $-0.75_{-0.15}^{+0.28}$ & $-0.79_{-0.70}^{+0.96}$ & $-0.10_{-0.38}^{+0.42}%
$\\
&  &  &  &  &  & \\
\multicolumn{7}{c}{\textbf{Dataset PP +OHD + BAO}}\\
$\mathbf{\Lambda}$\textbf{CDM} & $68.9_{-1.5}^{+1.5}$ & $0.311_{-0.012}%
^{+0.012}$ & $146.0_{-3.1}^{+3.1}$ & $-1$ & $0$ & $0$\\
$w$\textbf{CDM} & $68.1_{-1.6}^{+1.6}$ & $0.296_{-0.016}^{+0.016}$ &
$146.8_{-3.2}^{+3.2}$ & $-0.912_{-0.054}^{+0.054}$ & $0$ & $0~$\\
\textbf{Coexistence} & $68.1_{-1.7}^{+1.7}$ & $0.165_{-0.150}^{+0.052}$ &
$146.9_{-3.4}^{+3.4}$ & $-0.78_{-0.10}^{+0.24}$ & $-0.55_{-0.55}^{+0.96}$ &
$-0.08_{-0.26}^{+0.39}$\\
&  &  &  &  &  & \\
\multicolumn{7}{c}{\textbf{Dataset U3 +OHD + BAO}}\\
$\mathbf{\Lambda}$\textbf{CDM} & $68.8_{-1.6}^{+1.6}$ & $0.312_{-0.013}%
^{+0.013}$ & $146.3_{-3.5}^{+3.1}$ & $-1$ & $0$ & $0$\\
$w$\textbf{CDM} & $67.2_{-1.7}^{+1.7}$ & $0.297_{-0.016}^{+0.016}$ &
$146.8_{-3.3}^{+3.3}$ & $-0.858_{-0.065}^{+0.065}$ & $0$ & $0$\\
\textbf{Coexistence} & $67.0_{-1.7}^{+1.7}$ & $<0.113$ & $146.8_{-3.3}^{+3.3}$
& $-0.662_{-0.065}^{+0.21}$ & $-0.82_{-0.74}^{1.0}$ & $-0.03_{-0.36}^{+0.48}%
$\\
&  &  &  &  &  & \\
\multicolumn{7}{c}{\textbf{Dataset DESD +OHD + BAO}}\\
$\mathbf{\Lambda}$\textbf{CDM} & \thinspace$68.4_{-1.7}^{+1.7}$ &
$0.314_{-0.012}^{+0.0095}$ & $147.0_{-3.5}^{+3.5}$ & $-1$ & $0$ & $0$\\
$w$\textbf{CDM} & $67.8_{-1.7}^{+1.7}$ & $0.297_{-0.015}^{+0.013}$ &
$147.2_{-3.4}^{+3.4}$ & $-0.905_{-0.048}^{+0.048}$ & $0$ & $0$\\
\textbf{Coexistence} & $67.8_{-1.6}^{+1.6}$ & \thinspace$<0.141$ &
$147.3_{-3.4}^{+3.4}$ & $-0.707_{-0.062}^{+0.200}$ & $-0.85_{-0.75}^{+1.0}$ &
$-0.13_{-0.30}^{+0.45}$\\\hline\hline
\end{tabular}
\label{table1}%
%TCIMACRO{\TeXButton{E}{\end{table}}}%
%BeginExpansion
\end{table}%
%EndExpansion
\begin{figure}[h!]
\centering\includegraphics[width=1\textwidth]{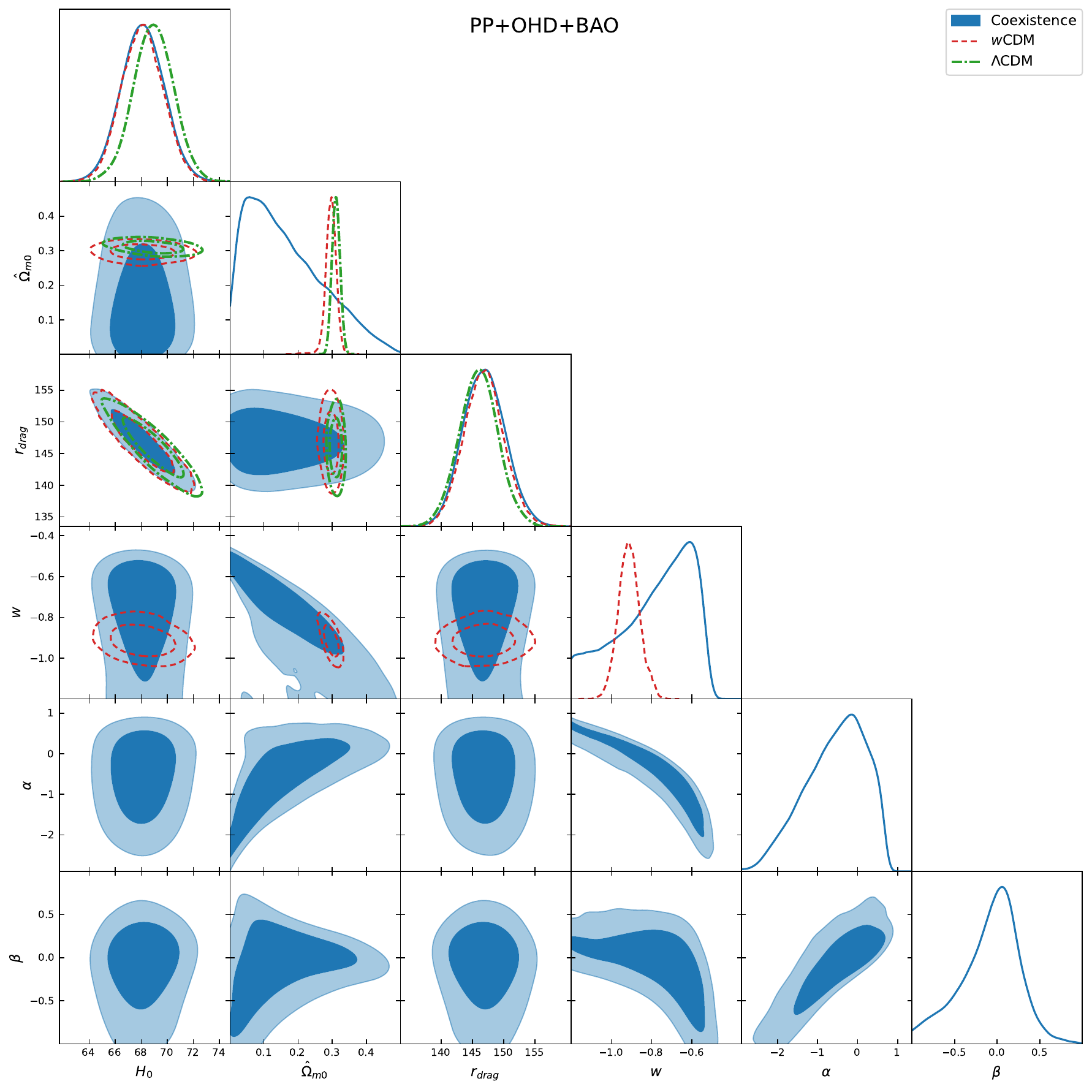}\caption{Dataset
PP + OHD + BAO: confidence space for the posterior parameters for the
coexistence Interacting, $w$CDM and $\Lambda$CDM models.}%
\label{fig1}%
\end{figure}

\begin{figure}[h!]
\centering\includegraphics[width=1\textwidth]{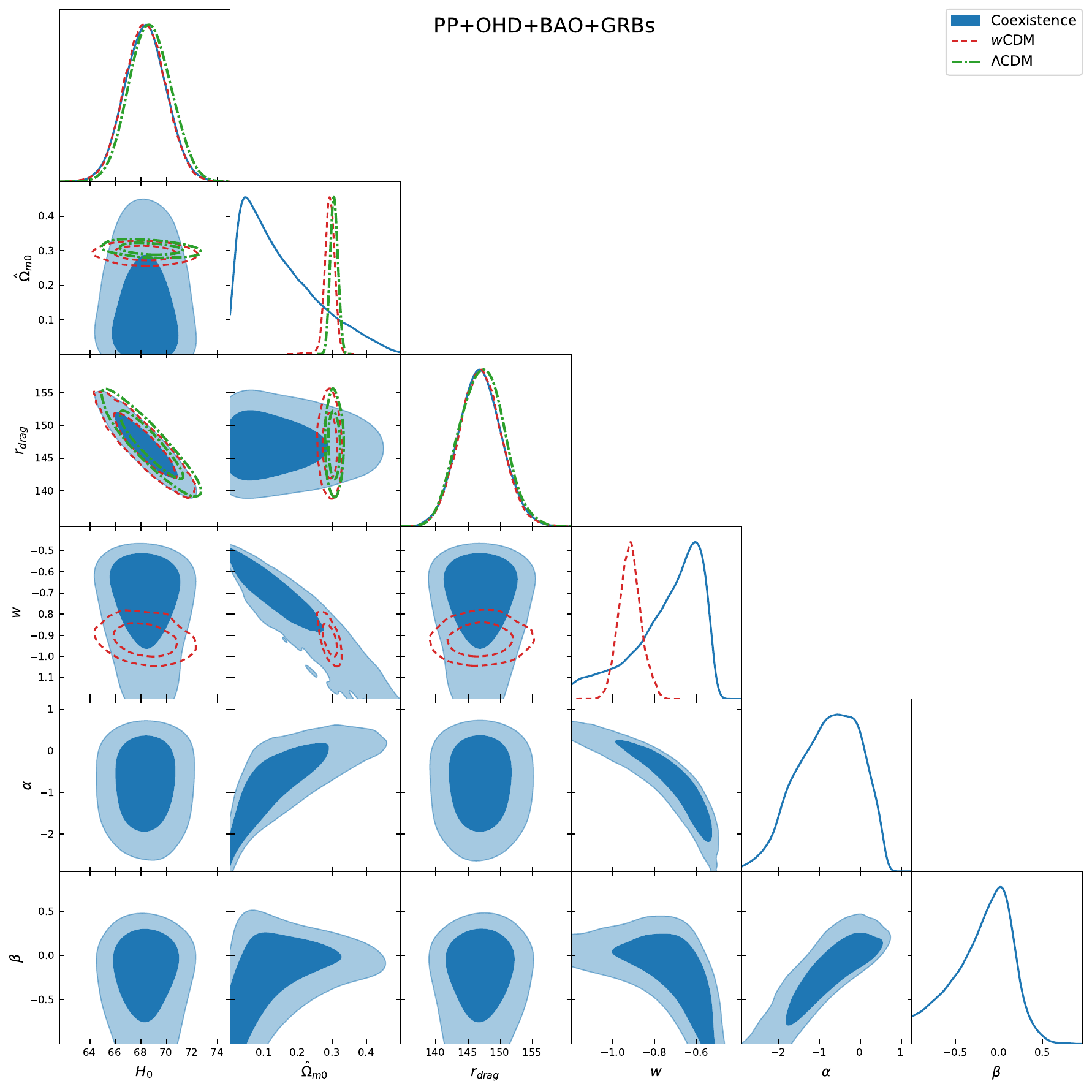}\caption{Dataset
U3 + OHD + BAO + GRBs: confidence space for the posterior parameters for the
coexistence Interacting, $w$CDM and $\Lambda$CDM models.}%
\label{fig2}%
\end{figure}

\begin{figure}[h!]
\centering\includegraphics[width=1\textwidth]{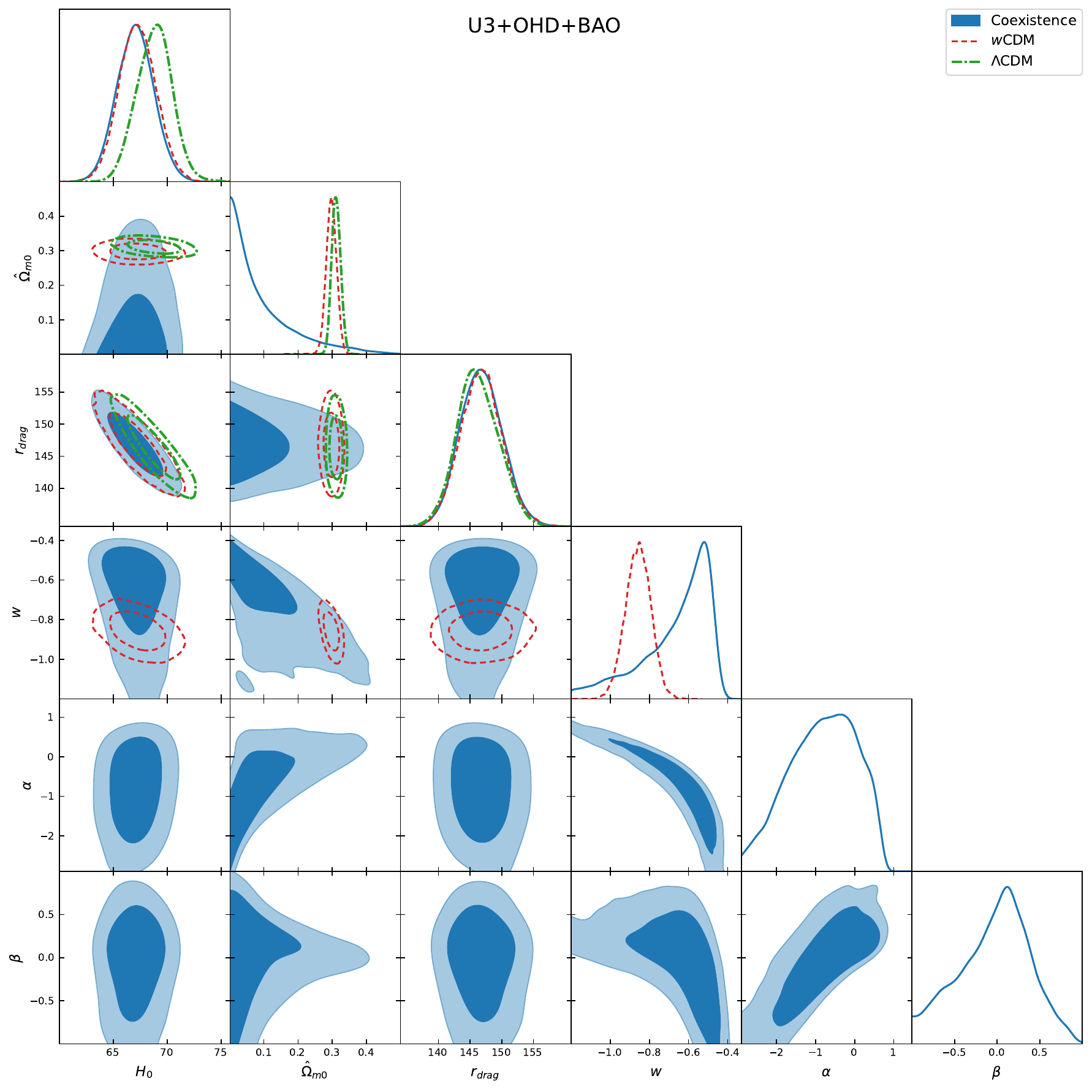}\caption{Dataset
U3 + OHD + BAO: confidence space for the posterior parameters for the
coexistence Interacting, $w$CDM and $\Lambda$CDM models.}%
\label{fig3}%
\end{figure}

\begin{figure}[h!]
\centering\includegraphics[width=1\textwidth]{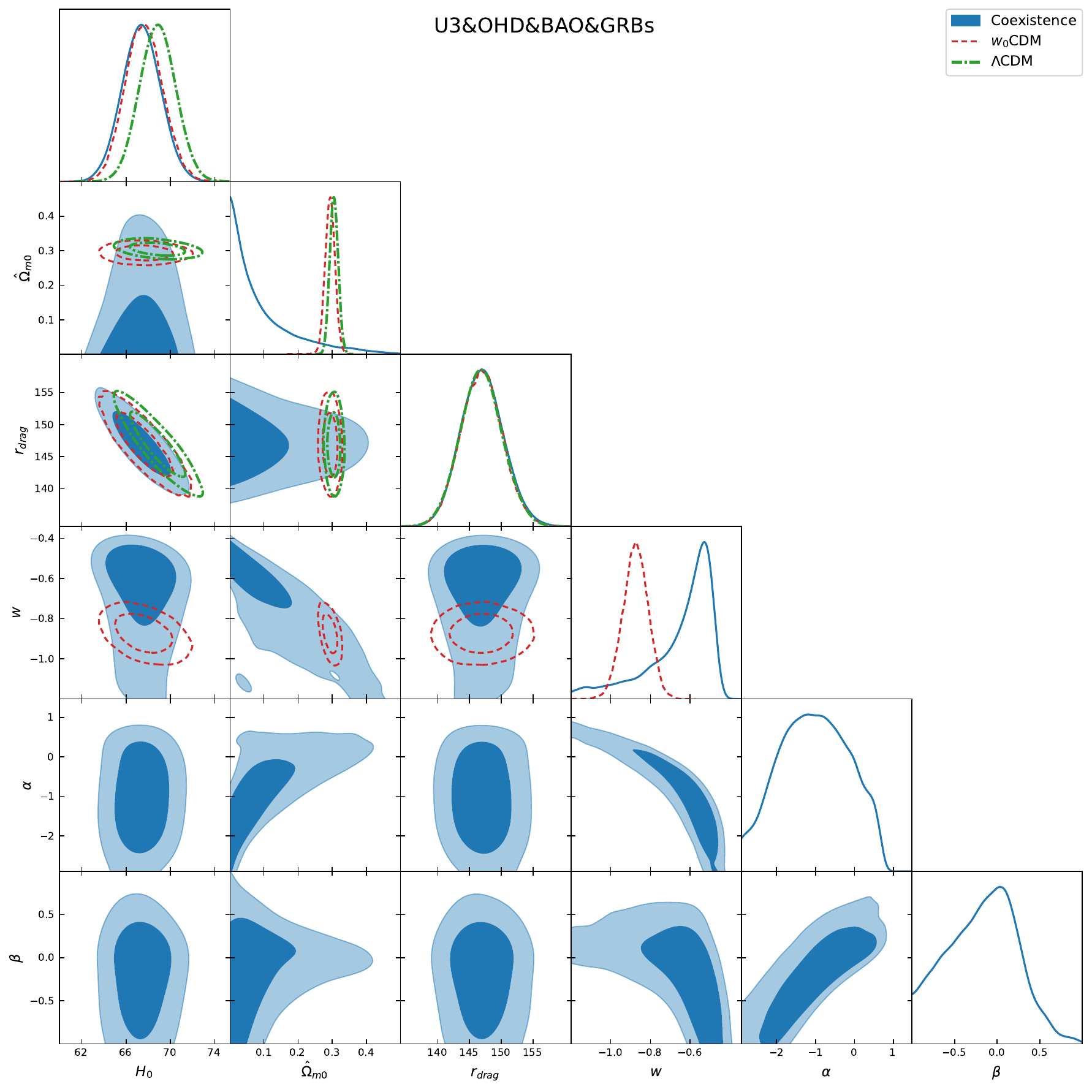}\caption{Dataset
U3 + OHD + BAO + GRBs: confidence space for the posterior parameters for the
coexistence Interacting, $w$CDM and $\Lambda$CDM models.}%
\label{fig4}%
\end{figure}

\begin{figure}[h!]
\centering\includegraphics[width=1\textwidth]{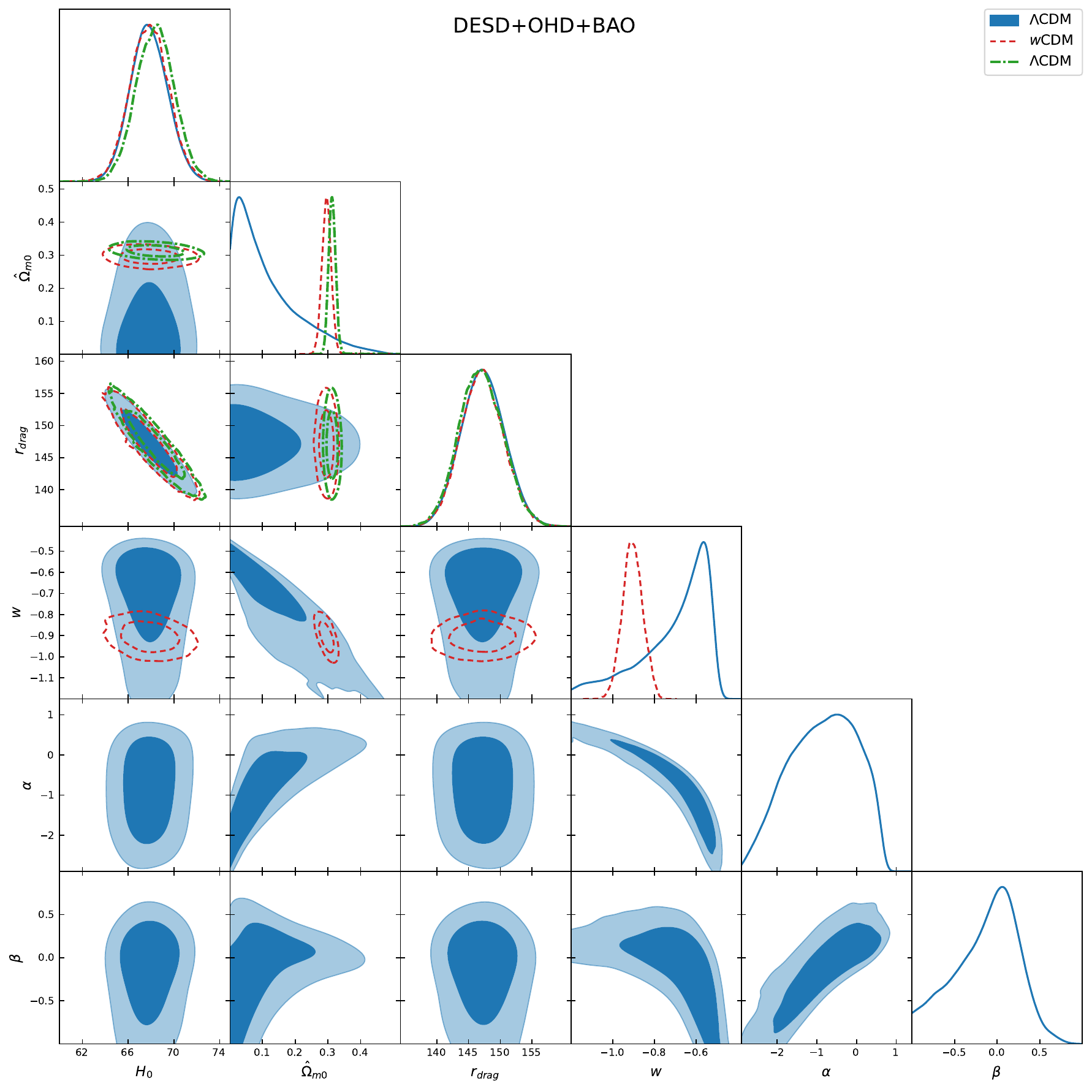}\caption{Dataset
DESD + OHD + BAO + GRBs: confidence space for the posterior parameters for the
coexistence Interacting, $w$CDM and $\Lambda$CDM models.}%
\label{fig5}%
\end{figure}

\begin{figure}[h!]
\centering\includegraphics[width=1\textwidth]{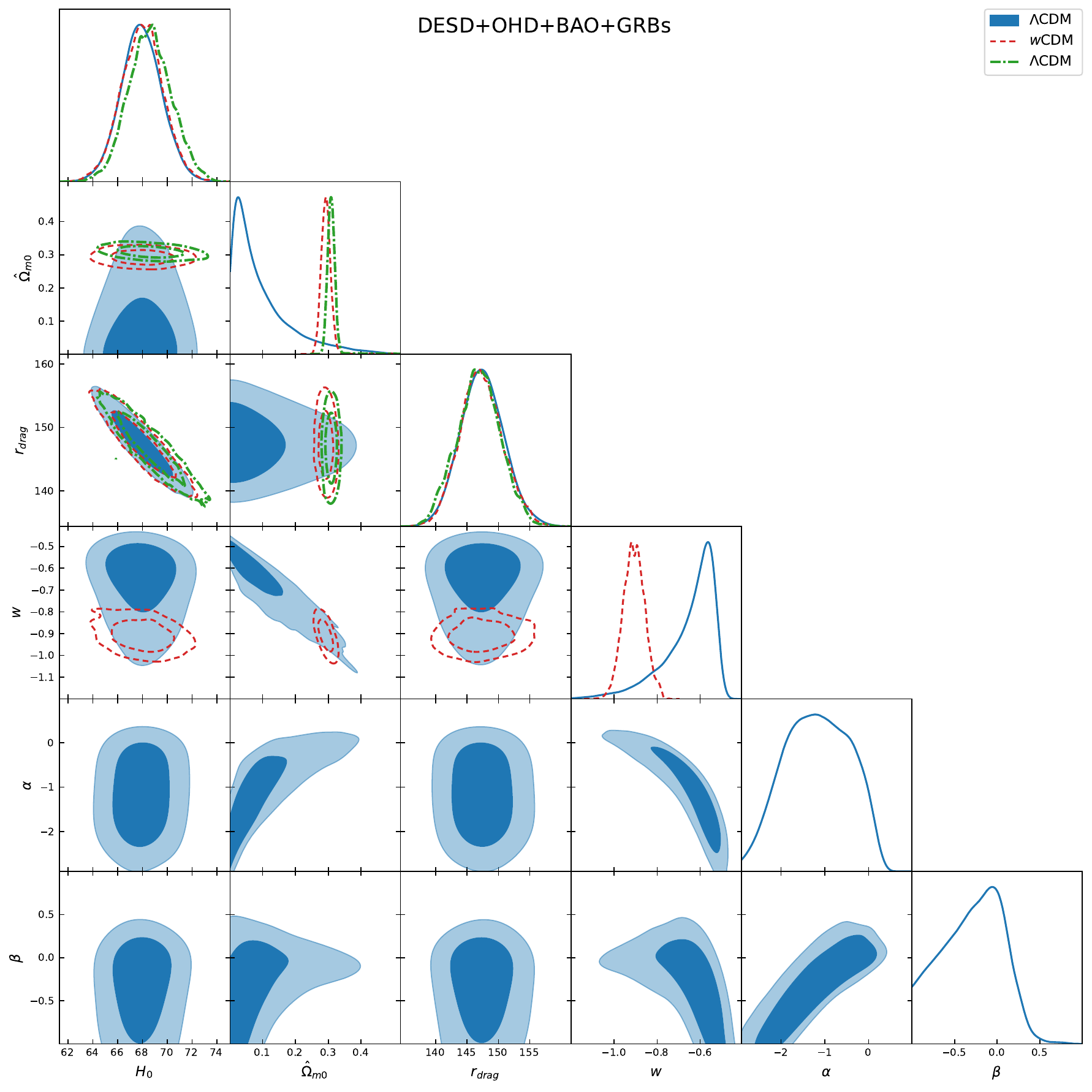}\caption{Dataset
DESD + OHD + BAO + GRBs: confidence space for the posterior parameters for the
coexistence Interacting, $w$CDM and $\Lambda$CDM models.}%
\label{fig6}%
\end{figure}

%TCIMACRO{\TeXButton{B}{\begin{table}[tbp] \centering}}%
%BeginExpansion
\begin{table}[tbp] \centering
%EndExpansion
\caption{Posterior Parameters for the interacting model which describes coexistence (2/2).}%
\begin{tabular}
[c]{ccccccc}\hline\hline
& $\mathbf{H}_{0}$ & $\mathbf{\hat{\Omega}}_{m0}$ & $\mathbf{r}_{drag}$ &
$\mathbf{w}_{d}$ & $\alpha$ & $\beta$\\
\multicolumn{7}{c}{\textbf{Dataset OHD + BAO + GRBs}}\\
$\mathbf{\Lambda}$\textbf{CDM} & $69.9_{-1.6}^{+1.6}$ & $0.290_{-0.014}%
^{+0.012}$ & $146.5_{-0.014}^{+0.012}$ & $-1$ & $0$ & $0$\\
$w$\textbf{CDM} & $71.1_{-2.4}^{2.4}$ & $0.287_{-0.014}^{+0.013}$ &
$146.9_{-3.3}^{+3.3}$ & $<-0.98$ & $0$ & $0$\\
\textbf{Coexistence} & $71.2_{-3.3}^{+2.8}$ & \thinspace$<0.176$ &
$146.9_{-3.3}^{+3.3}$ & $-0.86_{-0.27}^{+0.17}$ & $-1.17_{-0.77}^{+0.87}$ &
$-0.34_{-3.1}^{+0.31}$\\
&  &  &  &  &  & \\
\multicolumn{7}{c}{\textbf{Dataset PP +OHD + BAO + GRBs}}\\
$\mathbf{\Lambda}$\textbf{CDM} & $68.7_{-1.6}^{+1.6}$ & $0.306_{-0.011}%
^{+0.011}$ & $147.1_{-3.4}^{+3.4}$ & $-1$ & $0$ & $0$\\
$w$\textbf{CDM} & $68.3_{-1.6}^{+1.6}$ & $0.292_{-0.015}^{+0.015}$ &
$146.9_{-3.3}^{+3.3}$ & $-0.918_{-0.055}^{+0.049}$ & $0$ & $0~$\\
\textbf{Coexistence} & $68.3_{-1.6}^{+1.6}$ & $0.151_{-0.140}^{+0.043}$ &
$146.9_{-3.3}^{3.3}$ & $-0.740_{-0.076}^{+0.200}$ & $-0.78_{-0.65}^{+0.89}$ &
$-0.19_{-0.25}^{+0.40}$\\
&  &  &  &  &  & \\
\multicolumn{7}{c}{\textbf{Dataset U3 +OHD + BAO + GRBs}}\\
$\mathbf{\Lambda}$\textbf{CDM} & $68.9_{-1.6}^{+1.6}$ & $0.305_{-0.012}%
^{+0.012}$ & $146.9_{-3.3}^{+3.3}$ & $-1$ & $0$ & $0$\\
$w$\textbf{CDM} & $67.6_{-1.7}^{+1.7}$ & $0.294_{-0.015}^{+0.015}$ &
$146.9_{-3.3}^{+3.3}$ & $-0.875_{-0.064}^{+0.064}$ & $0$ & $0$\\
\textbf{Coexistence} & $67.4_{-1.7}^{+1.7}$ & $<0.110$ & $147.0_{-3.4}^{+3.4}$
& $-0.650_{-0.044}^{+0.190}$ & $-1.02_{-0.40}^{+0.45}$ & $-0.17_{-0.40}%
^{+0.45}~$\\
&  &  &  &  &  & \\
\multicolumn{7}{c}{\textbf{Dataset DESD +OHD + BAO + GRBs}}\\
$\mathbf{\Lambda}$\textbf{CDM} & $68.6_{-1.8}^{+1.8}$ & $0.311_{-0.013}%
^{+0.009}$ & $146.9_{-3.5}^{+3.5}$ & $-1$ & $0$ & $0$\\
$w$\textbf{CDM} & $68.0_{-1.7}^{+1.7}$ & $0.294_{-0.16}^{+0.12}$ &
$147.3_{-3.4}^{+3.4}$ & $-0.908_{-0.050}^{+0.050}$ & $0$ & $0$\\
\textbf{Coexistence} & $67.9_{-1.9}^{+1.9}$ & $0.099_{-0.096}^{+0.015}$ &
$147.4_{-3.5}^{+3.5}$ & $-0.650_{-0.096}^{+0.14}$ & $-1.16_{-0.73}^{+0.73}$ &
$-0.29_{-0.32}^{+0.43}$\\\hline\hline
\end{tabular}
\label{table2}%
%TCIMACRO{\TeXButton{E}{\end{table}}}%
%BeginExpansion
\end{table}%
%EndExpansion

\begin{itemize}
\item[-]OHD + BAO: The coexistence interacting model fits this dataset, the mean
posterior values and $1\sigma$ errors, $H_{0}=68.1_{-2.8}^{+2.3}$%
,~$\hat{\Omega}_{m0}<0.175$, $r_{drag}=146.9_{-3.3}^{+3.3}$, $w_{d}%
=-0.75_{-0.15}^{+0.28}$, $\alpha=-0.79_{-0.70}^{+0.96}$ and $\beta
=-0.10_{-0.38}^{+0.42}$. We observe that the $w$CDM and the $\Lambda$CDM
models are within the $1\sigma$ parametric space of the $\alpha$ and$~\beta$
coupling constants. The coexistence model fits the data in a better way in
comparison to the other two models, with smaller values for the$~\chi_{\min
}^{2}$, that is, $\chi_{C}^{2}-\chi_{\Lambda}^{2}=-1.71$ and $\chi_{C}%
^{2}-\chi_{w\mbox{CDM}}^{2}=-1.65$. However, due to the different number of
degrees of freedom, according to Akaike's scale, the dataset provides weak
evidence in favor of the $\Lambda$CDM.

\item[-] PP +OHD + BAO: The introduction of the PP catalogue provides the
posterior parameters $H_{0}=68.1_{-1.7}^{+1.7}$,~$\hat{\Omega}_{m0}%
=0.165_{-0.150}^{+0.052}$, $r_{drag}=146.9_{-3.4}^{+3.4}$, $w_{d}%
=-0.78_{-0.10}^{+0.24}$, $\alpha=-0.55_{-0.55}^{+0.96}$ and $\beta
=-0.08_{-0.26}^{+0.39},$ where again the two models of reference models are
within the $1\sigma$ parametric space of the $\alpha$ and $~\beta$ coupling
constants.. The comparison of the statistical parameters shows provides $\chi_{C}^{2}-\chi_{\Lambda}^{2}=-3.66$ and $\chi_{C}^{2}-\chi_{w\mbox{CDM}}%
^{2}=-0.92$, which means that the Coexistence model fits this dataset in a better way, however, due to the different number of free parameters, the
analysis shows a weak preference of this dataset to the $\Lambda$CDM and
$w$CDM models.

\item[-] U3 +OHD + BAO:\textbf{ }When we introduce the U3 catalogue for the
supernova data we find $H_{0}=67.0_{-1.7}^{+1.7}$,~$\hat{\Omega}_{m0}<0.113$,
$r_{drag}=146.8_{-3.3}^{+3.3}$, $w_{d}=-0.662_{-0.065}^{+0.21}$,
$\alpha=-0.82_{-0.74}^{1.0}$ and $\beta=-0.03_{-0.36}^{+0.48},$where now only
the $w$CDM is recovered is within the $1\sigma$ parametric space of the
$\alpha$ and$~\beta$ coupling constants. For the statistical parameters, we
calculate that $\chi_{C}^{2}-\chi_{\Lambda}^{2}=-8.22$ and $\chi_{C}^{2}%
-\chi_{w\mbox{CDM}}^{2}=-3.52$, which means that this dataset provides a weak
evidence in favor of the coexistence model, in contrast to the $\Lambda$CDM,
while coexistence interaction and the $w$CDM are statistically equivalent.

\item[-] DESD +OHD + BAO: The DESD catalogue for the supernova reveals the best
fit parameters $H_{0}=67.8_{-1.6}^{+1.6}$,~$\hat{\Omega}_{m0}$\thinspace
$<0.141$, $r_{drag}=147.3_{-3.4}^{+3.4}$, $w_{d}=-0.707_{-0.062}^{+0.200}$,
$\alpha=-0.85_{-0.75}^{+1.0}$ and $\beta=-0.13_{-0.30}^{+0.45}$, \ where now
only the only the $w$CDM is recovered is within the $1\sigma$ parametric
space of the $\alpha$ and$~\beta$ coupling constants. The comparison of the
statistical parameters shows that the coexistence model fits the data better
from the other models $\chi_{C}^{2}-\chi_{\Lambda}^{2}=-4.13$ and $\chi
_{C}^{2}-\chi_{w\mbox{CDM}}^{2}=-2.39.$ Nevertheless, from Akaike's scale, we conclude that the models are statistically equivalent. 

\item[-]OHD + BAO + GRBs: The introduction of the GRB data in theOHD + BAO
dataset modify the posterior space for the free parameters as $H_{0}%
=71.2_{-3.3}^{+2.8}$,~$\hat{\Omega}_{m0}<0.176$, $r_{drag}=146.9_{-3.3}%
^{+3.3}$, $w_{d}=-0.86_{-0.27}^{+0.17}$, $\alpha=-1.17_{-0.77}^{+0.87}$ and
$\beta=-0.34_{-3.1}^{+0.31}$. Where now the values $\alpha=\beta=0$ are not
within the $1\sigma$ region. Nevertheless, the coexistence model fits almost in a similar way as the two references model, that is, $\chi_{C}%
^{2}-\chi_{\Lambda}^{2}=-0.7$ and $\chi_{C}^{2}-\chi_{w\mbox{CDM}}^{2}=-0.21$,
while there exists a weak support in favor of the $\Lambda$CDM.

\item[-] PP +OHD + BAO + GRBs: For this dataset, we derive the posterior space
$H_{0}=68.3_{-1.6}^{+1.6}$,~$\hat{\Omega}_{m0}=0.151_{-0.140}^{+0.043}$,
$r_{drag}=146.9_{-3.3}^{3.3}$, $w_{d}=-0.740_{-0.076}^{+0.200}$,
$\alpha=-0.78_{-0.65}^{+0.89}$ and $\beta=-0.19_{-0.25}^{+0.40}$, where only
the $w$CDM is recovered within the $1\sigma$ regime of the $\alpha$
and$~\beta$ coupling constants.. The coexistence interaction and the $w%
$CDM\ fit the data in a similar way, but both fits better in comparison to the
$\Lambda$CDM, i.e. $\chi_{C}^{2}-\chi_{\Lambda}^{2}=-3.36$ and $\chi_{C}%
^{2}-\chi_{w\mbox{CDM}}^{2}=-0.78$. However, the application of the AIC does not
lead to any supportive conclusion for the coexistence model.

\item[-] U3 +OHD + BAO + GRBs: For the use of the U3 catalogue with the GRBs we
find $H_{0}=67.4_{-1.7}^{+1.7}$,~$\hat{\Omega}_{m0}<0.110$, $r_{drag}%
=147.0_{-3.4}^{+3.4}$, $w_{d}=-0.650_{-0.044}^{+0.190}$, $\alpha
=-1.02_{-0.40}^{+0.45}$ and $\beta=-0.17_{-0.40}^{+0.45}$, where now none of the reference models are recovered within the $1\sigma$ regime. The comparison
of the statistical parameters reads $\chi_{C}^{2}-\chi_{\Lambda}^{2}=-7.52$
and $\chi_{C}^{2}-\chi_{w\mbox{CDM}}^{2}=-3.85$, from where make the same
conclusions as before.

\item[-] DESD +OHD + BAO + GRBs: Finally, from the DESD catalogue we find the
best fit variables $H_{0}=67.9_{-1.9}^{+1.9}$,~$\hat{\Omega}_{m0}%
=0.099_{-0.096}^{+0.015}$, $r_{drag}=147.4_{-3.5}^{+3.5}$, $w_{d}%
=-0.650_{-0.096}^{+0.14}$, $\alpha=-1.16_{-0.73}^{+0.73}$ and $\beta
=-0.29_{-0.32}^{+0.43}$, where now none of the reference models are recovered
within the $1\sigma$ regime. The comparison of the statistical parameters
reads $\chi_{C}^{2}-\chi_{\Lambda}^{2}=-6.95$ and $\chi_{C}^{2}-\chi
_{w\mbox{CDM}}^{2}=-3.52$, from where we conclude that the three models similarly fit the data. 
\end{itemize}

From this analysis, we found that the coexistence model fits all the datasets in a better way than do the $\Lambda$CDM and $w$CDM models. However, the introduction of the GRBs on these datasets does not introduce any statistically significant result, except for the moment that within the
$1\sigma$ regime on the $w$CDM is recovered. Therefore, we can infer that
for large redshifts the coexistence interacting model and the $w$CDM have a similar behaviour.

We also performed the same analysis without the three new OHD data points. We
obtained essentially the same values for the free parameters and reached the
same conclusions. This indicates that the new measurements do not introduce
any tension in the models and appear to be consistent with the previously
known behaviour of the Hubble function.

Moreover, except for the Dataset OHD + BAO + GRBs, all other constraints
indicate that the parameter $\beta=0$, is within the $1\sigma$ of the
confidence space, which is the limit where the compartmentalization interacting model is recovered. Comparing the obtained results with those of \cite{vanderWesthuizen:2025iam} we can conclude that the two interacting
models are statistically equivalent.

Finally, in Figs. \ref{fig1}, \ref{fig2}, \ref{fig3}, \ref{fig4}, \ref{fig5}
and \ref{fig6} we present the confidence space of the free parameters for the
four datasets PP +OHD + BAO, PP +OHD + BAO + GRBs, U3 +OHD + BAO, U3 +OHD + BAO + GRBs,
DESD +OHD + BAO,  and DESD +OHD + BAO, respectively.%

\begin{table}[tbp] \centering
%EndExpansion
\caption{Comparison of the statistical parameters (1/2).}%
\begin{tabular}
[c]{ccccc}\hline\hline
& $\mathbf{\chi}_{C}^{2}\mathbf{-\chi}_{\Lambda}^{2}$ & $\mathbf{AIC}%
_{M}\mathbf{-AIC}_{\Lambda}$ & $\mathbf{\chi}_{C}^{2}\mathbf{-\chi}_{w%
CDM}^{2}$ & $\mathbf{AIC}_{C}\mathbf{-AIC}_{w\mbox{CDM}}$\\\hline
\multicolumn{5}{c}{\textbf{Dataset OHD + BAO}}\\
$\mathbf{\Lambda}$\textbf{CDM} & $0$ & $0$ & $0.06$ & $-1.94$\\
$w$\textbf{CDM} & $-0.06$ & $1.94$ & $0$ & $0$\\
\textbf{Coexistence} & $-1.71$ & $4.29$ & $-1.65$ & $2.35$\\
&  &  &  & \\
\multicolumn{5}{c}{\textbf{Dataset PP +OHD + BAO}}\\
$\mathbf{\Lambda}$\textbf{CDM} & $0$ & $0$ & $2.74$ & $0.74$\\
$w$\textbf{CDM} & $-2.74$ & $-0.74$ & $0$ & $0$\\
\textbf{Coexistence} & $-3.66$ & $2.34$ & $-0.92$ & $3.08$\\
&  &  &  & \\
\multicolumn{5}{c}{\textbf{Dataset U3 +OHD + BAO}}\\
$\mathbf{\Lambda}$\textbf{CDM} & $0$ & $0$ & $4.70$ & $2.70$\\
$w$\textbf{CDM} & $-4.70$ & $-2.70$ & $0$ & $0$\\
\textbf{Coexistence} & $-8.22$ & $-2.22$ & $-3.52$ & $0.48$\\
&  &  &  & \\
\multicolumn{5}{c}{\textbf{Dataset DESD +OHD + BAO}}\\
$\mathbf{\Lambda}$\textbf{CDM} & $0$ & $0$ & $\,3.48$ & $1.48$\\
$w$\textbf{CDM} & $-3.48$ & $-1.48$ & $0$ & $0$\\
\textbf{Coexistence} & $-4.13$ & $1.87$ & $-2.39$ & $1.61$\\\hline\hline
\end{tabular}
\label{table3}%
%TCIMACRO{\TeXButton{E}{\end{table}}}%
%BeginExpansion
\end{table}%
%EndExpansion
\bigskip%
%TCIMACRO{\TeXButton{B}{\begin{table}[tbp] \centering}}%
%BeginExpansion
\begin{table}[tbp] \centering
%EndExpansion
\caption{Comparison of the statistical parameters (2/2).}%
\begin{tabular}
[c]{ccccc}\hline\hline
& $\mathbf{\chi}_{C}^{2}\mathbf{-\chi}_{\Lambda}^{2}$ & $\mathbf{AIC}%
_{M}\mathbf{-AIC}_{\Lambda}$ & $\mathbf{\chi}_{C}^{2}\mathbf{-\chi}_{w%
CDM}^{2}$ & $\mathbf{AIC}_{C}\mathbf{-AIC}_{w\mbox{CDM}}$\\
\multicolumn{5}{c}{\textbf{Dataset OHD + BAO + GRBs}}\\
$\mathbf{\Lambda}$\textbf{CDM} & $0$ & $0$ & $0.48$ & $-1.52$\\
$w$\textbf{CDM} & $-0.48$ & $1.52$ & $0$ & $0$\\
\textbf{Coexistence} & $-0.7$ & $5.3$ & $-0.21$ & $3.79$\\
\multicolumn{5}{c}{\textbf{Dataset PP +OHD + BAO + GRBs}}\\
$\mathbf{\Lambda}$\textbf{CDM} & \thinspace\thinspace$0$ & $0$ & $-2.58$ &
$0.58$\\
$w$\textbf{CDM} & $-2.58$ & \thinspace\thinspace$-0.58$ & $0$ & $0$\\
\textbf{Coexistence} & $-3.36$ & $2.64$ & $-0.78$ & $3.22$\\
\multicolumn{5}{c}{\textbf{Dataset U3 +OHD + BAO + GRBs}}\\
$\mathbf{\Lambda}$\textbf{CDM} & $0$ & $0$ & $3.67$ & $1.67$\\
$w$\textbf{CDM} & $-3.67$ & $-1.67$ & $0$ & $0$\\
\textbf{Coexistence} & $-7.52$ & $-1.52$ & $-3.85$ & $0.15$\\
&  &  &  & \\
\multicolumn{5}{c}{\textbf{Dataset DESD +OHD + BAO + GRBs}}\\
$\mathbf{\Lambda}$\textbf{CDM} & $0$ & $0$ & $3.43$ & $1.43$\\
$w$\textbf{CDM} & $-3.43$ & $-1.43$ & $0$ & $0$\\
\textbf{Coexistence} & $-6.95$ & $-0.94$ & $-3.52$ & $0.48$\\\hline\hline
\end{tabular}
\label{table4}%
%TCIMACRO{\TeXButton{E}{\end{table}}}%
%BeginExpansion
\end{table}%
%EndExpansion

\section{Conclusions}

\label{sec5}

The models that describe interaction within the dark sector of the universe
allow us to unlock key insights into the behavior of dark matter and dark
energy. These models provide a framework to understand the impact of the dark
sector on cosmic evolution. They also help us to explore the dynamical
behaviour of the dark energy and to address the cosmological tensions.

Although these models are mostly phenomenological in nature, dark energy-dark
matter interacting terms can emerge in fundamental theoretical frameworks,
such as in Weyl Integrable Spacetime, in the Jordan frame, and others.

In this work, we followed a phenomenological approach and consider an
interaction in which the dynamical behavior of the physical parameters is
governed by a Lotka--Volterra system. This framework can lead to solutions
that describe coexistence within the dark sector.

To examine the physical viability of this interacting scenario, we performed a
series of observational tests by constraining the free parameters of the model
using late-time cosmological data. Specifically, we considered different
combinations of the following datasets: Cosmic Chronometers, DESI DR2 BAO
measurements, Supernovae from the PantheonPlus, Union 3.0, DES-Dovekie
catalogues, and Gamma-Ray Bursts

We also performed the same tests for the $\Lambda$CDM and the $w$CDM
models, which we considered as reference models. We found that, for all the
different tests we carried out, the interacting model fits the data better
than the reference models. However, due to the different numbers of degrees of
freedom, we made use of the AIC to determine which model is preferred by each
dataset. According to Akaike's scale, the coexistence model is always
statistically equivalent to the $w$CDM model, while the combination of the
BAO data with the Union3.0 catalogue poses a challenge to the $\Lambda$CDM
scenario, as the data show a weak preference in favor of the interacting model.

\textbf{From this study, we conclude that the coexistence interaction is physically
viable and can serve as a potential mechanism for describing the dynamical
behaviour of dark energy. However, we have not addressed the implications of
this model for the current cosmological tensions, such as the $H_{0}%
$--tension. A proper examination requires incorporating CMB data into the
analysis \cite{Li:2026xaz,Toda:2024fgv}, which we plan to pursue in future work.}

Moreover, we intend to investigate whether there exists an underlying theoretical mechanism capable of producing a dynamical evolution similar to
that was obtained in this phenomenological model.

\begin{acknowledgments}
AP \& GL thank AA and the Centre for Space Research at the North-West University for the hospitality
extended during the completion of this work. AP \&  GL acknowledge support from
FONDECYT Grant 1240514 and from VRIDT through Resolución VRIDT No.
096/2022 and Resolución VRIDT No. 098/2022.
\end{acknowledgments}

%\bibliographystyle{unsrt}
%\bibliography{references}
%\bibliographystyle{plain}
\bibliography{biblio}

\end{document}